\documentclass[emulateapj]{emulateapj}
\begin{document}
\shorttitle{Hybrid Mechanism}
\shortauthors{Currie, T.}
\title{Hybrid Mechanisms for Gas/Ice Giant Planet Formation}
\author{Thayne Currie}
\affil{Department of Physics and Astronomy \& UCLA Center for Astrobiology, University of California-Los Angeles, Los Angeles, CA 90095}
\email{currie@astro.ucla.edu}
\begin{abstract}
The effects of gas pressure gradients on the motion of solid grains in the solar nebula substantially
enhances the efficiency of forming protoplanetary cores in the standard core accretion model in 
'hybrid' scenarios for gas/ice giant planet formation.  Such a scenario is enhanced core
 accretion which
results from Epstein-drag induced inward radial migration of mm-sized grains and subsequent particle 
subdisk gravitational instability needed to build up
a population of 1 km planetesimals.  Solid/gas ratios can be enhanced by nearly $\sim 
10\times$ over those in Minimum Mass Solar Nebula (MMSN)
in the outer solar nebula (a $>$ 20 AU), increasing the oligarchic core masses and decreasing 
formation timescales for protoplanetary cores.  A 10 $M_{\oplus}$ core can form on 
$\sim 10^{6}-10^{7}$ year timescales at 15 - 25 AU compared to $\sim 10^{8}$ years in the standard model,
alleviating the major problem plaguing the core accretion model for gas/ice giant planet formation.
\end{abstract}
\section{Introduction}
The main challenges facing the standard core accretion model for planet formation
are finding a mechanism for building up a population of planetesimals and forming 
sufficiently massive proto-Uranus and proto-Neptune cores before the end of the 
oligarchic growth stage and before the solar nebula dissipates on $10^{7}$ yr timescales.

  In the core accretion model, planetesimals are built up from approximately $\mu$m-sized 
grains to km-sized bodies by collisional sticking.  Two major problems
face collisionally sticking these bodies together.
First, growing particles from centimeter through kilometer sizes is problematic.  While
many have argued for the coagulation of grains based on some requisite 'coagulation velocity'
(e.g. Weidenschilling 1997) such velocities are often much higher than those below which sticking 
might occur based on microgravity experiments such as Blum \& Muench (1993).  More recently, Colwell (2003)
showed that at least in some conditions cm-sized grain coagulation can occur, but this
occurs for relative velocities below $\sim 12 cm/s$ and greater relative velocities are
encountered in either a turbulent or laminar disk (see Youdin \& Shu 2002, hereafter YS02, and references
therein).
An operative sticking mechanism also remains in question: both known 
solid state sticking forces and the bodies' self gravity are arguably too weak (YS02). 
Second, even 
if such sticking were possible, the timescale for meter-sized bodies to spiral in to
 the Sun (by gas drag) is short (Weidenschilling 1977), $10^{2}$ years at 1 AU, such that
planetesimal formation in all but the outermost regions of the solar nebula might not 
proceed fast enough. One can circumvent this problem if a particle subdisk gravitational
 instability (GI) can very quickly build up a population of $1 km$ planetesimals directly 
(Goldreich \& Ward, 1973; YS02).  

However, formation timescale problems still 
remain for ice giants Uranus and Neptune.  For a MMSN distribution, the oligarchic
growth stage does not yield Neptune-mass cores (for a $>$ 15 AU), so that a $15 M_{\oplus}$ core must
 be formed by colliding together sub Earth-mass oligarchs.  Levison \& Stewart (2001) showed that 
the embryos' mutual perturbations result in a large number of ejections (not mergers) of embryos
in the outer solar system, so that the formation timescale for a $10-15 M_{\oplus}$ proto-Neptune core
is prohibitively long.  The two most widely explored attempts at a solution are positing a massive
($>$ $6\times$ MMSN) disk during the planet formation epoch (Goldreich et al., 2004, hereafter GO04) or
 that ice giant cores were gravitationally scattered to their present orbits
after brisker formation in the trans-Saturnian region (Thommes et al. 2002; 
Thommes et al. 2003).  However, these ideas, while plausible, introduce very case-specific assumptions
about the initial state of the solar nebula or its dynamical history.   

The purpose of this paper is to show that another mechanism for forming ice giant cores within the 
core accretion model exists that is more generally applicable to protoplanetary disks.  Specifically,
 one may be able to form Uranus and Neptune-mass cores in situ and on
solar nebula dissipation timescales from what is currently the
most plausible mechanism for inducing particle subdisk GI necessary for planetesimal formation.  
In Section 2 \& 3 we first review this planetesimal formation mechanism, called here the Sekiya-Youdin gravitational 
instability model, and then argue that it could aid ice giant formation in proceeding quickly.  Next, 
in Section 4 we describe our recalculation of grain pileups in this model originally done by YS02 and 
Youdin \& Chiang (hereafter YC04), develop a simple model for calculating protoplanetary core masses 
at the end of oligarchy and core formation timescales.  In Section 5 we present our the results of our core mass 
and formation timescale calculations, showing how the $\Sigma_{p}$, solid body surface density, profile
resulting from this planetesimal formation mechanism alters these values substantially.  In
the discussion section we put this model in a historical context, summarize it, and describe what research within this
model should be done further.  Finally, we include an appendix describing the protoplanetary disk conditions
necessary to allow GI.

\section{Planetesimal Formation from Sekiya-Youdin GI}
Unless one assumes a hitherto unknown efficient sticking mechanism for growing $1 cm- 1 km$ bodies 
one likely has to posit some sort of particle layer GI to build
up km-sized planetesimals.  Planetesimal formation by GI, as developed by Goldreich \& Ward
(1973), was long considered problematic, though.  The chief reason for this was that as particles settle towards the disk 
midplane the disk becomes sufficiently vertically stratified that Kelvin-Helmholtz turbulence
develops.  Turbulent eddies resulting from the Kelvin-Helmholtz instability induce particle random velocities
that are too high to allow GI, and equilibrium vertical profiles for particles result in spatial densities that are
at least an order of magnitude too low (Weidenschilling 1980; Cuzzi, Dobrovolskis, \& Champney 1993 
; Weidenschilling 1995). 

Sekiya et al. (1998), however, found that if the ratio of dust to gas surface densities was sufficiently enhanced
over standard solar values, turbulence induced by Kelvin-Helmholtz vertical shear will not be sufficient to stir all
the solids: the rest will be able to precipitate towards the disk midplane and induce GI.  YS02
 and YC04 provide a likely mechanism for achieving this enhancement by pileup of grains from gas 
drag-induced migration, avoiding the main difficulty with GI from Goldreich \& Ward (1973).  We now describe 
this mechanism following YS02 and YC04.

Because gas is more sensitive to pressure gradients, it orbits at slightly sub-Keplerian velocities where the 
relative velocity difference between gas and dust is proportional to $\eta=-{(\partial P_{g}/\partial 
ln a)}/{2\rho_{g}v_{k}^{2}}\sim ({c_{g}}/{v_{k}})^{2}$.  The dust then experiences a headwind, and thus a 
drag force.  For a stopping time of a particle moving relative to the gas of $\tau_{stop} = \rho_{p} s /
\rho_{g}c_{g}$, and gas and particle
densities of $\rho_{g}$ and $\rho_{p}$ respectively and particle sizes $s$, the inward particle flux through the disk is $f=f_{ep}+f_{turb}$, 
 migration due to Epstein drag is $f_{ep} =\rho_{p}(\frac{\rho_{g}}{\rho})^{2} v_{ep,ind}=
\rho_{p}({\rho_{g}}/{\rho})^{2} 2\eta\Omega \tau_{stop} v_{k}$ and turbulent stresses is $f_{turb}
\sim\rho_{p}v_{ep,ind}$ $\frac{\rho_{g}}{\rho} \frac{\partial}{\partial z}({\rho_{p}z}/{\rho})$ (whose effect
is small: see YC04).  This inward particle flux from Epstein drag-induced migration results in grain pileups
if the flux decreases with decreasing stellocentric distance $a$.  Explicitly, whether such a condition is
satisfied depends on the radial profiles for solid and gas surface densities $\Sigma_{p}$ and $\Sigma_{g}$
as well as the temperature profile.  

In a MMSN model, one starts with solid body gas profiles of $\Sigma_{p} \propto a^{-n}$ and $\Sigma_{g} \propto
a^{-p}$ and a temperature profile of $T \propto a^{-q}$.  The migration rate due to Epstein drag is 
\begin {equation}
v_{a}\sim v_{ep,ind} = 2\frac{\rho_{g}}{\rho}\eta \tau_{stop}\Omega ^{2}a.
\end {equation}
As $v_{a}$ depends on $\eta$ and $t_{stop}$ which in turn depend on T and $\Sigma_{g}$.  Isolating the
radial power law dependence of T and $\Sigma_{g}$ on $a$ one finds that the drift rate dependence goes as
$v_{a}\propto a^{d}$, where
\begin{equation}
d=p-q+1/2 .
\end{equation}
The mass accretion rate is $\propto \Sigma_{p}av_{a}\propto a^{E}$ where
$E= d-n+1$ or $E = \frac {3}{2} + p -n -q$ for grain migration in the Epstein drag regime .  If E $>$ 0 initially 
then the accretion rate increases with stellocentric distance, resulting in
particle pileups and thus $\Sigma_{p}$ enhancement. 
Epstein drag then results in grain pileups from the particle flux's sensitivity
to the gas density and temperature profiles.

Once enhancements of $\Sigma_{p}/\Sigma_{g}$ from migration are sufficiently high such that the 
solid particle surface density, $\Sigma_{p}$, rises above some critical value $\Sigma_{pc}$
 then particle subdisk GI (and thus planetesimal formation) commences
from dust in excess of $\Sigma_{pc}$.  Epstein drag-induced migration brings about particle layer GI
 because pileup of solid grains from migration can raise $\Sigma_{p}$ above $\Sigma_{pc}$, overcoming
Kelvin-Helmholtz shear instabilities. Particle layer gravitational instabilities are allowed
  once the solid to gas ratio is enhanced by $5-20\times$ (see YC04).  Garaud \& Lin (2004) find a similar
criteria for GI using a more sophisticated two-fluid treatment.

\section{Enhanced Efficiency of Protoplanet Core Accretion from Sekiya-Youdin GI}
YS02 and YC04 showed how Epstein drag-induced grain migration can procure planetesimal formation 
but didn't explore the effect that such migration has on the efficiency of planet formation.  
 Epstein drag-induced migration of grains, as shown by YS02 \& YC04, yields substantially
 different solid body surface density profiles from an MMSN profile, enhancing $\Sigma_{p}$.  Since
the criteria for GI (and planetesimal formation) is a $5 - 20x$ enhancement of $\Sigma_{p}$, 
the mass of planetesimals available form which to form protoplanets should greatly increase
within the dust subdisk as the mass from which such planetesimals are formed is locally enhanced by 
importing material from the outer solar nebula via migration.

  As the masses of post-oligarchic cores as
 well as protoplanet formation and inelastic collisional damping and collision timescales depend strongly
 on $\Sigma_{p}$ a seemingly simple enhancement of $\Sigma_{p}$ by about an order of magnitude could have
 very important implications for the efficiency of planet formation in regions with such enhancement. 
 The effect wasn't obvious from YS02 and YC04 since both papers did not include solid enhancement from water
 ice condensation in their initial $\Sigma_{p}$ profiles.  For example, model H in YC04 then triggers GI at 
an outer boundary of only $\sim$ 2 AU while model Af triggers GI at an inner boundary with 
 an enhancement of only $\sim 1.5 - 2\times \Sigma_{p,MMSN}$ at 15-25 AU when one compares the resulting
$\Sigma_{p}$ profile with an MMSN profile including the water ice enhancement.  However, if one includes water ice
 condensation in $\Sigma_{p,init}$ then pileup should proceed faster further away from the Sun.  Planetesimal
 formation could then be triggered earlier and the resulting $\Sigma_{p}$ profile could influence planet formation 
efficiency in the outer solar system (10-30 AU). 

\section{Calculations}
 We then are motivated to redo the YS02 calculations (equations 22-30) for the radial migration of mm-sized grains
 but include the solid enhancement for $a > 3 AU$ due to the water ice condensation.  We also estimate
core masses expected at the end of the oligarchic growth stage as well as formation timescales for 
$10 M_{\oplus}$ cores.  
\subsection {Recalculation of Grain Radial Drift/Pileup}
Following YS02 we assume a 
population of uniformly-sized particles comprising the solid body column $\Sigma_{p}$.  Its time-dependent profile is
\begin{equation}
\frac{\partial \Sigma_{p}}{\partial t} = v_{a}\frac{\partial \Sigma_{p}}{\partial a}+\frac{\Sigma_{p}}{a}\frac{\partial}{\partial a}(av_{a}),
\end{equation}
which has an analytical solution:
\begin{equation}
\Sigma_{p}(a,t)=\Sigma_{o}a^{-1-d} a^{d+1-n}_{i}(a,t)
\end{equation}
where $d = 1/2 p - q$ and $a_{i}(a,t)$ is the initial location of a particle 
now at a after time given by Eq. 26 of YS02:
\begin{equation}
a_{i}(a,t)=a(1-(d-1)\frac{v_{a}t}{a} )^{-1/(d-1)}.
\end{equation}
 
\subsection {Oligarchic Core Masses and Core Formation Timescales}
We then estimate oligarchic core masses and core formation timescales in the standard MMSN model and
the model modified for migration of mm-sized grains.  Oligarchic growth ends in once 
oligarchs' velocity excitations from mutual viscous stirring overtakes their damping rates which
result from dynamical friction with smaller planetesimals.  This condition is met once 
the "surface density" of protoplanets, $\Sigma_{M}=M_{pro}/2\pi a \delta a$ where $\delta a$ is the feeding zone size,
 becomes comparable to the residual surface density of smaller bodies such that $\Sigma_{M} \sim \Sigma_
{m}(t)$ (GO04).  Assuming that little mass is yet lost from the system this condition can be related
to the initial surface density of solids:
\begin{equation}
 \Sigma_{M} +\Sigma_{m}(t) = \Sigma_{p}
\end{equation}
 Solving for $M_{pro}$ then yields the protoplanet mass at the end of oligarchy ($M_{olig}$):
\begin{equation}
M_{olig} \sim 0.04 (b/10)^{1.5}(\Sigma_{p}/10 gcm^{-2})^{1.5}(a/1AU)^{3} M_{\oplus}.
\end{equation}
After oligarchy ends dynamical interactions between oligarchs results in ejection of many of them,
not mergers (Levison \& Stewart 2001).

For calculating the core formation timescale we generally follow Ida \& Lin (2004) and Kokubo \&
Ida (2002) with some modifications depending on the size of accreted planetesimals gleaned from
Rafikov (2004).  From Kokubo \& Ida (2002), the mass accretion rate of a protoplanetary core depends
on $\Sigma_{p}$, the mass of the core at time t, the core's size $r_{p}$, its stellocentric distance
${a}$, and the velocity dispersion of accreted planetesimals:
\begin{equation}
\dot{M}=C\pi\Sigma_{p}\frac{2GMr_{p}}{<e^{2}>^{\frac{1}{2}}a^{2}\Omega},
\end{equation}
where $<e^{2}>^{\frac{1}{2}}$ is the rms eccentricity of planetesimals and C is a factor of order
unity.  The planetesimals' rms
eccentricity is found by equating the viscous stirring and gas drag timescales $T_{vs}=T_{gas}$
(Kokubo \& Ida (2002)).  The formation timescale is then given as $\tau_{c,acc}=\frac{M}{\dot{M}}$.
The approximate formation timescale for a core is then given by Ida \& Lin (2004):
%\begin{equation}
\begin{eqnarray}
\tau_{c,acc} \sim 1.2\times10^{5}(10 gcm^{-2}/\Sigma_{p})
(2400 gcm^{-2}/\Sigma_{g})^{0.4} \nonumber\\ \times(a/1AU)^{0.6}
(M_{c}/M_{\oplus})^{1/3}(m/10^{18}g)^{2/15} years,
\end{eqnarray}
%\end{equation}
where $m$ is the mean accreted planetesimal mass for a protoplanet in a core feeding zone size of $10 R_{Hill}$ 
($b = 10$).   The rate of accretion of $\sim100m - 10km$ planetesimals can be described by the above equation
and accretion is then said to be 'dispersion dominated'.
% The formation timescale
%increases with $m$ because higher mass planetesimals are excited to higher random velocities and eccencrities
% upon encounters. 
However, if the accreted planetesimals are small enough their relative velocities to cores are set
by differential shear within the Hill radius of the core.  In this case, the approach velocity of a planetesimal
to the core is $\sim \Omega R_{H}$ where $R_{H}$ is the core's Hill radius, and the vertical component of 
the planetesimal's velocity, $v_{z}$, small.  For a small enough $v_{z}$ the planetesimal disk becomes very thin
and the embryo can accrete the entire vertical column of planetesimals (Rafikov 2004).  This regime of accretion
is called 'shear-dominated accretion' as the shear from the planetesimal disk around a core sets that core's 
velocity.  Rafikov (2004) then shows that the formation timescale for protoplanetary cores accreting shear-
dominated planetesimals then is shorter.  We rewrite the timescale here, generalizing it to variables $\Sigma_{p}$,
 $\rho_{p}$, $M_{e}$, and $\chi$:
\begin{eqnarray}
\tau_{c,acc} \sim 2.36\times10^{4}\chi^{-1}(10 gcm^{-2}/\Sigma_{p})
\nonumber\\ \times(M_{e}/M_{\oplus})^{1/3}(\rho_{p}/1 gcm^{-3})^{1/6} years,
\end{eqnarray}
where $\chi$ is the fraction of planetesimals which are shear dominated and $\rho_{p}$ is their mass density.
 These isolated body equations for $M_{olig}$ and $\tau_{c,acc}$ should be valid as long as
 growth occurs prior to the end of oligarchy and on timescales comparable to or less than the complete 
dissipation of the gas disk. 
\subsection{Size of Accreted Planetesimals}
The size of accreted planetesimals is likely to be smaller than the $\sim 1km$ bodies formed after
GI.  Kilometer-sized objects inelastically collide and fragment if their collisional velocities exceed  
$Q_{D}$, the energy per gram needed to release half of the planetesimal's mass
upon collision.  For a 300 m projectile hitting a 1km target $Q_{D} \sim 4\times 10^{4} erg/g$ (Leinhardt
 \& Richardson 2002), where $Q_{D}$ decreases for mass ratios approaching unity: low $Q_{D}$'s are 
consistent with the standard low internal strength, 'rubble pile' model for icy $km$-sized bodies such 
as comets (e.g. Asphaug \& Benz 1996).  The random velocity needed to disrupt a 1km planetesimal is $\sim 
10 m/s$ for a comparably-sized projectile of $s = 500 m$.  This velocity is comparable to $v_{H}$ at
30 AU.  The velocity of collisions actually 
experienced between $\sim 1 km$-sized planetesimals near a protoplanetary core is set by the Hill velocity,
 $v_{H}$, because their random velocities are set by scattering encounters with the core (GO04, Kenyon \& 
Bromley 2004, and Rafikov 2004). Furthermore, random velocities after a scattering encounter are likely to
 exceed $v_{H}$ (see Rafikov 2004, equations 23, 27, 42, and 46) unless the bodies already have $s << 1 km$.
 This means that 1 km planetesimals scattered by protoplanetary cores should fragment
 upon collision with one another, an outcome confirmed by numerical simulations of planetesimal interactions
around cores (e.g. Kenyon \& Bromley 2004).  

This is important because, as was alluded to before, the size of planetesimals very fundamentally affects core
accretion rates.  Specifically, for small planetesimals accretion can be much more rapid when planetesimal 
accretion occurs in the aforementioned 'shear-dominated' regime.  We now describe the conditions for such
accretion as they are given in Rafikov (2004).

Whether planetesimal accretion is shear or dispersion dominated depends largely on values for a fiducial mass,
$M_{f}$ and a fiducial size, $r_{f}$ which relate the planetesimal mass and sizes to planetesimal velocities 
induced after embryo-planetesimal scattering events.  In particular, $M_{f} = M_{\odot}\eta^{3}\sim8\times10^24
(a/1AU)^{3/2} g$ and $r_{f}=0.2\frac{\Sigma_{g}\Omega a}{\rho_{s} cs}\sim 120 (a/1AU)^{-7/4} m$.  Also important
is a size, $r_{s}$, where the Reynolds number of the gas, $Re$, is $Re=Re_{b}=20$, corresponding to changes
in the drag coefficient experienced by a planetesimal moving through gas and where the planetesimal's velocity
dispersion equals its velocity difference with the gas: $v=\delta v_{g}$(Rafikov 2004).  This size is given as
\begin{equation}
 r_{s}= \frac{\lambda R_{eb}}{3}\frac{\Omega a}{c_{s}} \approx 2.5 (a/1AU)^{5/2} m,
\end{equation}
where $\lambda \sim 1.25(a/1AU)^{11/4} cm$: the molecular mean free path ($\Sigma_{g}$ is reduced by a factor of
1.25 compared to Rafikov (2004) affecting $r_{f}, \lambda$, and $r_{s}$).

As shown in equations (42) and (46) of Rafikov (2004), the velocity of a planetesimal relative to the Hill velocity, 
$v_{H}$, depends on the core mass, $M_{c}$, as well as on $M_{f}$, $\lambda$, $r_{f}$, and $r_{s}$.  If $v \le v_{H}$
then shear-dominated accretion sets in, and thus we can describe the boundary between dispersion and shear-dominated
accretion as it depends on these parameters.  For the Stokes drag regime the condition for shear-dominated accretion is
\begin{equation}
M_{c}\le (\frac{r_{s}}{r_{f}})^{3}(\frac{r_{f}}{s})^{6}
\end{equation}
and for the Epstein drag regime it is
\begin{equation}
M_{c}\le (\frac{r_{s}}{\lambda})^{3}(\frac{r_{f}}{s})^{3}
\end{equation}
Thus, after some time of planetesimal accretion, the planetesimals must collide frequently enough such that the population
is ground down, keeping $s$ small, so that shear dominated accretion may persist.

\subsection{Model Assumptions and Initial Conditions}

For the radial migration calculations we begin with the following distributions of solids and gas. 
 We set $\Sigma_{o}=\epsilon\times 10 g cm^{-2}$ where $\epsilon = 1$ interior to $3 AU$, and
$\epsilon=4.2$ exterior.  The initial gas \&  solid surface
densities are given as $\Sigma_{g}=2400\times(a/1AU)^{-p} gcm^{-2}$ and $\Sigma_{p}=\Sigma_{o}\times(a/1AU)^{-n}$
and temperature profile given as $T=280\times(a/1AU)^{-q} K$ where p \& n = 1.5, and q = 0.5, corresponding to
Model H in YS02.  The disk initially extends out to 250 AU in this model, though the outer edge
of the solid component of the disk moves inward with time.  For this model, the drift rate can be given as
\begin{equation}
v_{a}\sim 3 (a/1 AU)^{1.5} \frac{10 \rho_{p} s}{g cm^{-2}} \frac {1 AU}{10^{6} yr}.
\end{equation}
We assume that the grains are comparable in size to the largest chondrules ($s = 1 mm$) to make direct comparisons
for the same run in YS02.  We do calculations for the formation timescale of a $10 M_{\oplus}$ core, one with the 
mean size of accreted planetesimals, $s$, set at 1 km, another with $s = 100m$ to take into account fragmentation, 
and $s=10m$ to further explore shear-dominated
accretion.  The density, $\rho_{p}$, is set to $3 gcm^{-3}$, representing an upper limit for planetesimals and ice-rock cores.  

\section{Results}
The solid particle migration run was stopped once one part of the disk had the requisite enhancement
to induce a particle layer gravitational instability according to YS02.  Figure 1 shows the time evolution
of $\Sigma_{p} (a,t)$ from an initial MMSN profile to the onset of particle subdisk GI.  When the bump
from water ice condensation is included in the surface density profile the time for the onset of GI
is $\sim 10\times$ shorter and occurs much further out at $\sim$ 27 AU than in the same model in YS02 and YC04,
yielding $\Sigma_{p}(a,t)/\Sigma_{p}(a,0)\sim 6-8\times$ from 20-25 AU and $\sim 9.15\times$ at the outer
 edge of the particle subdisk ($26.7 AU$) before GI is initiated at $t = 4.25\times 10^{4}$ yr.  In the 
'enhanced core accretion' scenario, the disk contains far more material from which to form
 protoplanets in precisely the regions where such formation has been most problematic for the standard core accretion
 model.  As the figure shows, this occurs because the initial distribution in $\Sigma_{p}$ is now 'squeezed' into a smaller 
surface area bounded roughly by the ice giant planet forming region due to Epstein drag-induced migration.

Figure 2 shows the masses of oligarchs for standard and migration-enhanced core accretion
 models (labeled Sekiya-Youdin migration).  For the latter, the oligarchic mass for Jupiter at 5.2 AU is $3.4 M_{\oplus}$, 
$21.1 M_{\oplus}$ for Uranus at 20 AU and $87 M_{\oplus}$ for Neptune at 25 AU, values much greater than in the standard
model, implying that Uranus and Neptune could have accreted most, if not all, of their mass ($15-17 M_{\oplus}$)by the
 end of the oligarchic growth phase.   The masses of embryos
at the end of oligarchy are large enough that additional, post-oligarchic growth can be negligible.  Dynamical
friction exerted by smaller bodies during oligarchic growth should dampen the embryos orbital eccentricities
inhibiting ejection.  Substantial post-oligarchic growth by coalescing is then unnecessary since cores have
already grown sufficiently massive during oligarchy.  Extra protoplanetary mass in the outer solar nebula
is ejected after oligarchy ends (GO04).

Figure 3 compares formation timescales for a $10 M_{\oplus}$ protoplanet core for the
standard core accretion model and the model utilizing the particle subdisk GI from Sekiya and Youdin's work
 to set $\Sigma_{p}$.  For the standard MMSN surface density 
profile the formation timescale for a $10 M_{\oplus}$ proto-Neptune core is $4.4\times10^{8}$
 and $1.75\times10^{8}$ years at 25 AU for $s = 1 km$
and $s = 100m$ respectively.  Since $\tau_{c,acc} $$>$$\tau_{dissip}$ and $M_{olig}$ is significantly less than
 $M_{neptune}$, the MMSN model will likely fail to produce Neptune-like planets on requisite timescales. 
  However, for the enhanced model the formation timescales are $5.5\times10^{7}$ and $2.2\times10^{7}$
 years at 25 AU for s = 1 km and s = 100m.  It then could be possible to form proto-Neptune nearly in
 situ on a timescale less than the disk dissipation timescale (10-30 Myr) for $s \le 100m$.  

Furthermore, a formation timescale of 20 Myr at 25 AU is most likely an overestimate as, using arguments from
Rafikov (2004), protoplanets here should grow a large fraction of their mass while the accreted
planetesimals are in the fast, shear-dominated regime.  Initially one starts with 1-10 km planetesimals where
the initial growth rate is fast.  However, once the core mass increases enough, planetesimal accretion becomes
dispersion dominated and growth occurs much more slowly. 
To get back to faster, shear-dominated accretion, the
initial population of planetesimals must drop in size enough such that $s << 1 km$.  Quantitatively, we refer
back to the equations (11) through (13).   At 25 AU $M_{f}$ and $r_{f}$, are $\sim 10^{27}g$ and $\sim 0.43 m$
respectively and $r_{s}$ is $\sim 7.8 km$.  Then, in the Stokes regime, dispersion-dominated accretion 
persists for $M_{e} \ge 3.8\times10^{19} g$ and $3.8\times10^{25}g$ for $s = 1 km$ and $100m$ the latter still
not an appreciable fraction of Neptune's current mass. 

 However, the $\sim 7-9x$ enhancement resulting from migration and pileup of mm-sized grains and subsequent GI helps
to bring one back into the shear-dominated accretion regime earlier and persist there for longer times by
the following argument.  This enhancement drives down the collision timescale by nearly an order of magnitude as
the collision rate is proportional to the number density of planetesimals and thus $\Sigma_{p}$
(e.g. Wetherhill \& Stewart 1993, equation A1).  This increases the dominance of collisions and collisional damping
 over gas damping and inspiral of solids  (the latter timescale is longer than gas damping by a factor of
 $1/(1.6\times10^{-3} a_{au}^{1/2})\sim125$).  If $\Sigma_{p}$ is enhanced we should expect more frequent collisions 
early on.  As the collisions tend to be disruptive at expected impact speeds of $v\ge v_{H}$, an enhancement in 
$\Sigma_{p}$ should result in planetesimal swarms of smaller mean sizes.  

The smaller sizes translate into planetesimal dynamics occurring in different drag regimes and longer times spent in
shear dominated planetesimal accretion.  Specifically, at 25 AU,
 planetesimals are in the Epstein regime for $s \le\lambda \sim 88m$. As this is slightly below an order of magnitude
reduction in the size of the initial swarm and collision rates between planetesimals are greater for enhanced $\Sigma_{p}$,
 it is likely that an appreciable fraction of accreted solids are in the 
Epstein drag regime and this regime is reached earlier than when a standard MMSN $\Sigma_{p}$ distribution
 is used.  One stays in the shear-dominated, Epstein drag regime as long as
 $M_{e}\le5\times10^{26}g$ and $6\times 10^{28}g$ for $s= 50m$ and $10m$, the latter being comparable to
the present rock/ice mass of Neptune. In the context of planetesimal accretion this means that cores can accrete
 planetesimals in the fast shear-dominated regime through a larger fraction of their total mass.  This has a drastic
 effect on the core formation timescale.  For shear-dominated accretion, we then find the formation timescale for a
 $10 M_{\oplus}$ core at 25 AU to be $\tau \sim 2.26\times 10^{5}
\chi^{-1} yr = 2.26\times 10^{6} yr$ if 10\% of planetesimals are shear dominated (shown in Figure 3) and $2.26
\times10^{5} yr$ if all of them are: both values are less than the solar nebula dissipation timescale, $\tau_
{dissip}\sim10^{7} yr$.  Formation timescales in the outer solar nebula are then sufficiently reduced and 
conditions are more likely to be met such that accelerated protoplanet growth persists through a sufficient
 fraction of proto-Neptune's final mass.

\section{Discussion}
The general idea in this paper that an enhancement of solid material in the Uranus \& Neptune regions of the solar 
nebula during the first 10 Myr is what led to rapid formation of ice giants is not new.  Various
incarnations of this idea have appeared in the last twenty years (e.g. Lissauer 1987; Pollack et al.1996; GO04) and 
even when such an enhancement was not considered the primary factor in forming Neptune on gas dissipation
 timescales, $\tau_{dissip}$, it often plays a supporting role in increasing a proto-Neptune's feeding zone
zone (e.g. Bryden et al. 2000).  These papers, however, usually assume such an enhancement of solids over MMSN
values results solely from the disk mass itself being much greater than MMSN values.  This paper relaxes that 
requirement, assuming a $\Sigma_{g}$ and initial $\Sigma_{p}$ profile comparable to MMSN values (total disk
 mass here is $\sim 0.05 M_{\odot}$ spread over 250 AU, out to 30 AU it is $\sim 0.02 M_{\odot}$), with very
 large enhancements of the solids achieved from grain migration and GI.
Thus, in this model it may be possible to get the same benefit you have from an initially massive disk (e.g. 
more solids from which to form planets, higher planetesimal collision rates, etc.) without the 'cost' of
having that disk mass be unreasonably large. 

That a global radial redistribution of solids influences some part of the planet formation process has also
been suggested previously.  Specifically, 
Stepinski \& Valageas (1997) and Kornet et al. (2001) posited that gas drag in a turbulent disk result in solid
 surface density distributions that differ from the gas surface density, providing enhancements in the inner 
regions of the solar nebula when compared to their initial distributions.  These papers,though, use a very
 simple treatment for grain
 coagulation for $1 cm - 10 km$ bodies, the size regime where a hierarchical sticking hypothesis seems to be
 highly problematic.  The effect any solid body redistribution has on the efficiency of post-planetesimal
 growth wasn't explored: rather, at least for Kornet et al. (2001) emphasis was placed on how different initial
conditions for the disk mass and radial extent explain the diversity of planetary systems via their differences
in grain redistribution.  It appears that grain redistribution can lead to not only a diversity of system
architectures but changes in the efficiency of planet formation.

The way in which planetesimal formation, particularly that of the GI model for planetesimal
 formation,influences the efficiency of planet formation has been hitherto underexplored and 
warrants further investigation.  A comparison between analytical estimates of core formation timescales and
numerical calculations assuming a $\Sigma_{p}$ distribution set by dust migration models such as YS02 should
 be made.  Furthermore, 
incorporating the 'shear-dominated' planetesimal accretion model into such calculations is needed as many papers
making pronouncements about the efficiency of protoplanetary core formation (e.g. Thommes et al. 2003) adopt
a prescription for the mass accretion rate similar to that for 'dispersion-dominated accretion' in this paper,
where a planetesimal's rms equilibrium eccentricity is set by balancing viscous stirring by the protoplanet 
with gas drag from the solar nebula.  As argued here and especially in Rafikov (2004), if the planetesimals
are small then accretion might proceed more rapidly.

If one considers the effect that Epstein drag-induced migration has on the distribution
of solids, the formation of cores in the outer solar nebula occurs on far shorter $\le 10^{7} yr$ timescales
 than in the standard core accretion model.  Assuming an initially massive disk or requiring 
substantial gravitational scattering of proto-Uranus and proto-Neptune cores, while still plausible, may be 
unnecessary for forming Uranus and Neptune in $\le 10^{7} years$.   Then 'hybrid' mechanisms such as this one
 resulting from Sekiya-Youdin particle subdisk GI could eliminate core formation timescale problems in the 
outer solar system.  Furthermore, the mechanism outlined in this scenario is generic if planetesimal formation 
proceeds by particle subdisk gravitational instability as described by YS02 \& YC04, operating regardless of
 whether others do.  Gas pressure gradients then play an indispensible role in forming the cores of gas/ice giant
 planets quickly, as they do in the rival disk instability model (see Boss et al. 2002, Haghighipour \& Boss
 2003), though in a slightly more indirect (and less obvious) way.  
\acknowledgements
The author would like to thank Andrew Youdin, Scott Kenyon, and Richard Nelson for extremely 
fruitful discussions and Nader Haghighipour, Brad Hansen, and the anonymous referee for suggestions
 that strengthened the arguments presented in this paper.  This research was 
supported in part by the NASA Astrobiology Institute.
\begin{appendix}
\centerline{DISK CONDITIONS FOR GRAIN SETTLING/GI}
It is important to ask whether 
the conditions for particle subdisk gravitational instability can be met.  Specifically, the main requirement
 for GI is that the disk must be sufficiently passive
to initially allow the dust layer to collapse to a thickness set by a balance between settling and Kelvin-
Helmholtz shear.  If viscous stirring is strong enough to prevent this amount of settling then the entire
GI mechanism for forming planetesimals becomes highly problematic.  We show here that 
the assumption that GI can occur is at least reasonable.

We first analyze previous attempts at constraining the equilibrium dust layer scale height achieved by balancing
settling with diffusivity generated by Richardson turbulence.  YC04 calculated
this criteria to be $\alpha \le 10^{-7} (a/1AU)^{1+p-q}$ or $\alpha \le 10^{-7}(a/1AU)^{2}$ for a MMSN model.
The coefficient $\alpha$ must then be less than $\sim 6.3\times10^{-5}$ for 
 $a = 25$ AU.  More 
stringent requirements can be found from Dubrulle et al. (1995).  From their equations 36 and
37, an equilibrium distribution of solids is written in terms of the ratio between the dust and
gas scale heights (h and H), the friction (stopping) time $\tau_{stop}$, and $\alpha$: 
\begin{equation}
\frac{h}{H}\sim \sqrt{{\alpha/\Omega\tau_{stop}}}.
\end{equation}
  The criteria for GI in terms of volume density is $\rho_{cr}\sim M_{\odot}/
a^{3}$ or $\sim 2.4\times10^{-11} g cm^{-3}$ or about a factor of $1000\times$ increase over its solar
abundance (and $\sim 100\times$ the gas density).  We then can rewrite the above condition as
$(\Sigma_{d}/100\Sigma_{g})^{2} \sim 10^{-6}\sim \alpha/(\tau_{stop}\Omega)$.
As $\tau_{stop}\Omega\sim10^{-4}(a/1AU)^{1.5}$ the criteria then becomes $\alpha \le 10^{-8}$ for 
grain sedimentation to a requisitely thin dust subdisk.  While estimates for $\alpha$ from molecular
viscosity are small enough ($\sim 10^{-12}$), values for $\alpha$ needed to explain mass accretion rates
onto T Tauri stars range from $10^{-2} - 10^{-4}$.

  Thus at first glance it seems as though conditions
 for GI include an implausibly low disk viscosity.  However, as pointed out by YC04, turbulence
need not be isotropic, and when $\alpha$ is resolved into components the criteria radially is given as
$\alpha_{r} \le \Omega t_{stop} \sim 10^{-4}(a/1AU)^{3/2}$ (from their equation B2).  The condition at 
25 AU is then that $\alpha_{r} \le 1.25\times10^{-2}$ which is reasonable.   The effective $\alpha_{r}$
 may differ substantially from the $\alpha_{z}$ if the viscosity source is more effective at mixing
 radially than vertically.  
 An example of a ratio between horizontal and vertical viscosities of order 
$10^{4}$ is cited by YC04 for Earth's oceans and atmosphere.       

A stronger argument can probably be made based on investigating disk structure
models in terms of whether a particular region is or is not viscously evolving,
especially considering the effect of dust grains on levels of disk viscosity. 
The disk may spend a substantial fraction of its lifetime accreting through only thin surface layers 
sandwiched between a quiescent, magnetically dead layer containing a large fraction of the total
 column of the disk material in so called 'layered disk' models (Gammie 1996).  The anomalous 
 viscosity then presumably tapers off and disappears beyond a certain gas column density into the disk
 a result consistent with MHD simulations of layered disks from Fleming \& Stone (2003). 
 Fromang et al. (2002) shows that for $\alpha = 10^{-2}$ and $\dot{M} \sim 10^{-8} M_{\odot}/yr$
the outer radius of the dead zone can be as large as ~100 AU and occupy 90 $\%$ of the vertical
 column density of the disk for metal fractions, $x_{M}$, slightly below cosmic abundances. 
  For $\alpha=10^{-3}$ a dead zone exists for all values
of the disk metal fraction, not just subsolar values. Furthermore, these 
calculations were done assuming that the grains do not affect the magnetic coupling to the disk, and the
authors specifically assumed that settling of the grains to the disk midplane had already occurred:
thus their ability to inhibit coupling of magnetic field lines to the gas would be irrelevant except for
a vanishingly thin dust layer at the midplane.  Fleming \& Stone (2003) also suggest that including the
 effects of grains would inhibit the ability of magnetic field lines to couple to the gas.

In absence of this assumption, Fromang et al. (2002) suggests that \textit{because} grains are effective at 
'scavenging' charge they may \textit{induce} a magnetically dead state within the subdisk they are contained, 
preventing turbulence from the anomalous source of viscosity from stirring them up and thus allowing 
sedimentation to the disk midplane if vertical mixing isn't particularly strong.  This
intuition now seems to have been confirmed numerically.
Recently,Nelson et al. (2005, in prep) has investigated the effect that dust grains have on the coupling
of magnetic field lines to the gas in the disk: thus their effect on the level of MRI
 turbulence.  This work suggests that dust grains are particularly efficient at scavenging
charge in the disk, and their inclusion leads to dead zone $\Sigma_{g}$'s that are much
larger than those in Fromang et al. (2002).  They find that in order to drive MRI turbulence
throughout the disk cross section at some distance, one needs to reduce the number density
of dust grains by a factor of $10^{8}$.   Accretion through the entire vertical column
of the disk would then seem to \textit{presume} that substantial grain growth and settling had
already occurred.  Clearly, one  
cannot at the moment make a strong case against the existence of conditions necessary for GI.  
\end{appendix}

\begin{figure}
    \centering
%   {\centering \resizebox*{0.9\textwidth}{!}{{\includegraphics{surdenevol.ps}}{\includegraphics{surdenevol.ps}}{\includegraphics{surdenevol.ps}}}\par}
%   {\centering \resizebox*{1.0\textwidth}{!}{{\includegraphics{surdenhyb.ps}}
%   {\includegraphics{surdenhyb.ps}}}\par}
   {\centering \resizebox*{0.5\textwidth}{!}{{\includegraphics{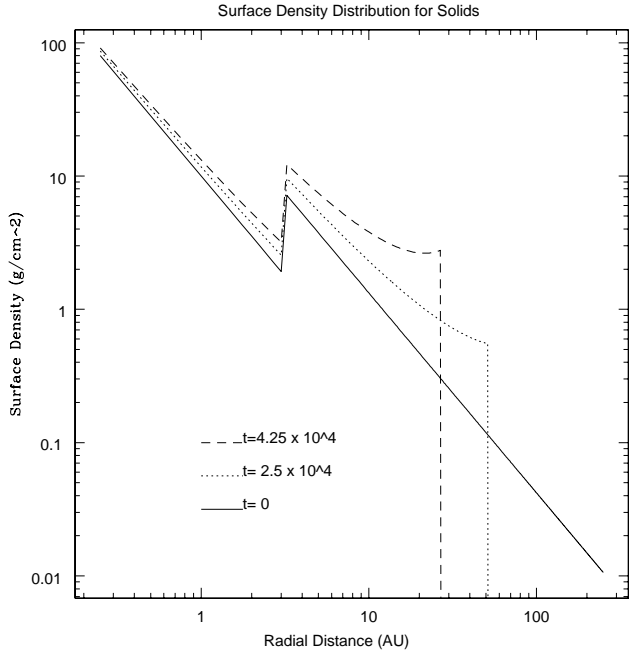}}}\par}
%   {\centering \resizebox*{0.5\textwidth}{!}{{{\includegraphics{surden.ps}}
%   {\includegraphics{enhhyb1.ps}}}\par}
%   {\centering \resizebox*{0.5\textwidth}{!}{{\includegraphics{olighyb.ps}}}\par}
%   {\includegraphics{surdenhyb.ps}}}\par}
%   {\centering \resizebox*{0.3\textwidth}{!}{{\includegraphics{surdenevol.ps}}}}
%   {\centering \resizebox*{0.4\textwidth}{!}{{\includegraphics{surdenevol.ps}}{\includegraphics{surdenevol.ps}}}\par}
%   \hspace{1in}%
%   {\centering \resizebox*{0.4\textwidth}{!}{\includegraphics{surdenevol.ps}}\par}
%   \hspace{1in}%
%   \centering
%   \includegraphics[width=0.5in]{surdenevol.ps}
   \caption{Evolution of the solid body surface density profile, $\Sigma_{p}$.  For this model (equivalent to
 Model H in YS02 \& YC04), Epstein drag 
experienced by mm-sized grains causes them to migrate, pile up, and induce particle subdisk GI to form
planetesimals.  This leaves a $\sim$ 8-9x enhancement in the outer regions of the particle
disk after $\sim 4.25\times 10^{4}$ yr, yielding a $\Sigma_{p}$ differing significantly from the standard
MMSN profile.}\label{figure 1}
\end{figure}
\begin{figure}
    \centering
%   {\centering \resizebox*{0.9\textwidth}{!}{{\includegraphics{surdenevol.ps}}{\includegraphics{surdenevol.ps}}{\includegraphics{surdenevol.ps}}}\par}
%   {\centering \resizebox*{1.0\textwidth}{!}{{\includegraphics{surdenhyb.ps}}
%   {\includegraphics{surdenhyb.ps}}}\par}
   {\centering \resizebox*{0.5\textwidth}{!}{{\includegraphics{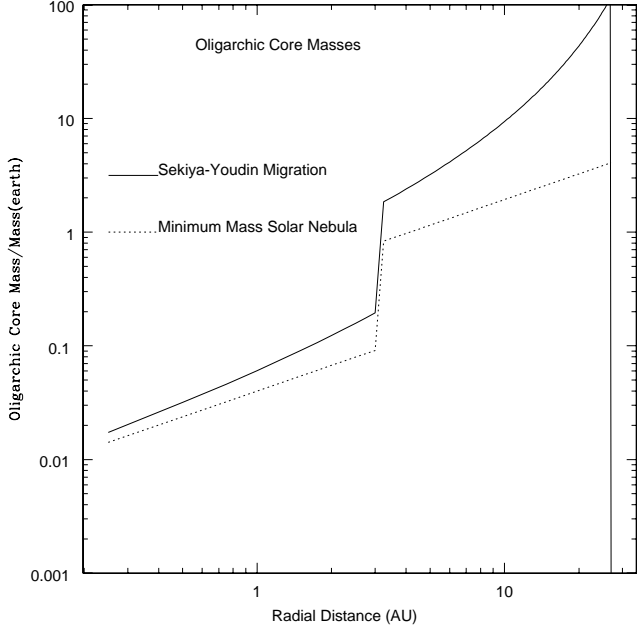}}}\par}
%   {\includegraphics{surdenhyb.ps}}}\par}
%   {\centering \resizebox*{0.3\textwidth}{!}{{\includegraphics{surdenevol.ps}}}}
%   {\centering \resizebox*{0.4\textwidth}{!}{{\includegraphics{surdenevol.ps}}{\includegraphics{surdenevol.ps}}}\par}
%   \hspace{1in}%
%   {\centering \resizebox*{0.4\textwidth}{!}{\includegraphics{surdenevol.ps}}\par}
%   \hspace{1in}%
%   \centering
%   \includegraphics[width=0.5in]{surdenevol.ps}
   \caption{Protoplanetary core masses at the end of oligarchic growth for the standard
core accretion model and the 'hybrid', migration-enhanced model.  $M_{olig}$ is greater than the predicted solid bodymasses for Uranus and Neptune at roughly their present orbits in the latter: all of their growth
could conceivably occur during oligarchy.}\label{figure 2}
\end{figure}
%\begin{figure}
%    \centering
%   {\centering \resizebox*{0.9\textwidth}{!}{{\includegraphics{surdenevol.ps}}{\includegraphics{surdenevol.ps}}{\includegraphics{surdenevol.ps}}}\par}
%   {\centering \resizebox*{1.0\textwidth}{!}{{\includegraphics{surdenhyb.ps}}
%   {\includegraphics{surdenhyb.ps}}}\par}
%   {\centering \resizebox*{0.5\textwidth}{!}{{\includegraphics{isohyb.ps}}}\par}
%   {\includegraphics{surdenhyb.ps}}}\par}
%   {\centering \resizebox*{0.3\textwidth}{!}{{\includegraphics{surdenevol.ps}}}}
%   {\centering \resizebox*{0.4\textwidth}{!}{{\includegraphics{surdenevol.ps}}{\includegraphics{surdenevol.ps}}}\par}
%   \hspace{1in}%
%   {\centering \resizebox*{0.4\textwidth}{!}{\includegraphics{surdenevol.ps}}\par}
%   \hspace{1in}%
%   \centering
%   \includegraphics[width=0.5in]{surdenevol.ps}
%   \caption{Isolation core mases for the standard and hybrid models.  The latter model
%enhances the amount of solid material out of which planets can form.  Then the planet formation
%process need not be as efficient as that in the standard model.  For example, the isolation core
%mass at Uranus' present orbit is $\sim$ 1.5-2x less than its current mass while that in the 
%hybrid model is over 3x its current mass.}  \label{figure 5}
%\end{figure}
\begin{figure}
    \centering
%   {\centering \resizebox*{0.9\textwidth}{!}{{\includegraphics{surdenevol.ps}}{\includegraphics{surdenevol.ps}}{\includegraphics{surdenevol.ps}}}\par}
   {\centering\resizebox*{0.5\textwidth}{!}{{\includegraphics{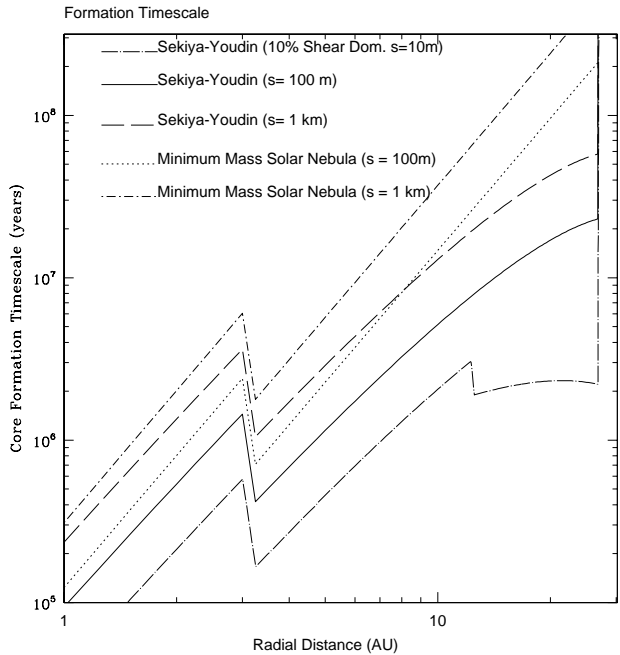}}}\par}
%   {\includegraphics{coretimea.ps}}}\par}
%   {\centering\resizebox*{0.5\textwidth}{!}{{\includegraphics{surdenhyb.ps}}}\par}
%   {\includegraphics{surdenhyb.ps}}}\par}
%   {\centering \resizebox*{0.3\textwidth}{!}{{\includegraphics{surdenevol.ps}}}}
%   {\centering \resizebox*{0.45\textwidth}{!}{{\includegraphics{surdenevol.ps}}{\includegraphics{surdenevol.ps}}}\par}
%   \hspace{1in}%
%   {\centering \resizebox*{0.4\textwidth}{!}{\includegraphics{surdenevol.ps}}\par}
%   \hspace{1in}%
%   \centering
%   \includegraphics[width=0.5in]{surdenevol.ps}
\caption{Formation timescales for $10 M_{\oplus}$ protoplanetary cores for 1km and
100m mean accreted planetesimal sizes.  The enhancement of $\Sigma_{p}$ after particle 
subdisk GI due to Epstein drag-induced migration reduces the core formation timescale by 
nearly 10x for dispersion-dominated planetesimals. $10 m$-sized planetesimals drop into 
the shear-dominated regime for $a$$>$$12.5 AU$ and dominate protoplanet accretion at these
distances such that $\tau_{acc} \approx 10^{6} yr$ to the edge of the planetesimal disk if
$10 \%$ of them are $10m$.
Thus, for $s $$<$$ 100m$ proto-Neptune can form on timescales $\tau_{acc} \approx \tau_{dissip}$,
 (the lifetime of the solar nebula) or less, even at 25 AU.} \label{figure 3}
\end{figure}

\end{document}